\documentclass[a4paper, 12pt]{article}

\usepackage{graphicx}
\usepackage{epstopdf}
\usepackage{amsmath}
\usepackage{subcaption}
\usepackage{caption}
\usepackage[round, sort]{natbib}
\usepackage{color}
\usepackage{bm}
\usepackage{authblk}
\usepackage{comment}
\usepackage[bottom]{footmisc}

\let\oldequation\equation
\let\endoldequation\endequation

\renewenvironment{equation}
  {\begin{footnotesize}\oldequation}
  {\endoldequation\end{footnotesize}}

\title{Impact of shear and dilational membrane viscosities on capsule shape and dynamics under shear flow: a theoretical study}
\author{P. Regazzi\thanks{Email address for correspondence: paul.regazzi@univ-amu.fr}\hspace{5pt} and M. Leonetti\thanks{Email address for correspondence: marc.leonetti@univ-amu.fr}}
\affil{Aix Marseille Univ, CNRS, CINaM, Marseille, France.}

\begin{document}
\maketitle
\begin{abstract}
In artificial (micro)capsules or cells such as red blood cells for example, the two-dimensional (2D) rheology of the membrane (or the shell) is viscoelastic and not only elastic. The deformation and inclination of such a capsule are theoretically studied under shear and extensional flows. The membrane mechanics is described by the Kelvin-Voigt model: the 2D dissipation with the Boussinesq-Scriven law associated to specific dilational and shear Boussinesq numbers $Bq_d$ and $Bq_d$ respectively while the elastic constitutive law is either Hookean, strain-softening (neoHookean) or strain-hardening (Skalak). The dynamics depends on the elastic capillary number $Ca$, the Poisson ratio (or $C$ in Skalak law), the viscosity ratio $\lambda$, the nonlinearities included in the constitutive law and both boussinesq numbers. Three dynamics are identified in the limit of weak deformations: a tank-treading motion in the limit of small elastic capillary number with a characteristic time depending on $\lambda$, $Bq_s$ and $Bq_d$, an oscillation without any elasticity and a damped oscillation leading to a tank-treading motion in the limit of large Boussinesq numbers. In all cases, both shear and dilational 2D dissipations contribute differently leading to non-trivial results. In the case of a Skalak law with $C=1$, $Bq=Bq_s=Bq_d$ and $Bq>>\lambda$, the deformations in the limits $Ca<<1$, $Bq=O(1)$ and  $Ca=O(1)$, $Bq>>1$ determined in the seminal work of \cite{barthes1985role} are recovered. 
\end{abstract}
\vfill
\section{Introduction}

The properties, abilities and sustainability of soft multiphasic systems depend on the mechanical and physico-chemical properties of their interfaces which bounded the system in the most simple cases such as core-layer droplets or are dispersed all over in more complex systems such as foams for example. Whatever the scale, these fields know a growing interest since several decades as first many applications require a multiphasic architecture made with soft materials, second core-layer to protect, transport and deliver molecules or molecular aggregates of vital necessity and third, core-layer can be considered as models of biological cells if a biomimetic two-dimensional viscoelastic rheology of the membrane is considered.\\

The core-layer soft microparticles are widespread unit elements in cosmetics, food, drug delivery, building materials. Although the core is a viscous droplet (or viscoelastic), the layer surrounding it varies greatly in nature, and the unique characteristics of this layer give the droplet its distinctive properties. In simplest examples, the layer is a fluid mono- or bi-layer of phospholipids, copolymers, biopolymers leading to vesicles and polymersomes (\cite{dionzou2016comparison,zhang2017polymersomes}) among other things. Their membrane mechanics is mainly governed by a resistance to bending and a local incompressibility leading to a rich zoology of shapes at thermal equilibrium (\cite{seifert1997configurations}), the ability of thin tubules to emerge (\cite{waugh2009membrane}) under local force (\cite{smith2004pulling}), spontaneous curvature (\cite{lipowsky2012spontaneous}), molecular activity (\cite{iyer2022non}) or an external field (\cite{waugh1982surface,shishkin2026metastability,boedec2013sedimentation,rangamani2013lipid}) and various fluid-structure configurations leading to subtle shapes and motions (\cite{young2026soft,chatkaew2009dynamics}). Vesicles are known to exhibit a high richness of spatiotemporal dynamics under linear and quadratic flow : \cite{biben2011three,deschamps2009dynamics,vlahovska2007dynamics,lyu2023swirling,kaoui2009red,agarwal2020stable}. The tank-treading motion arising under shear flow is the prototype at moderate viscosity ratio. Polymersomes differ from vesicles by their large shear membrane viscosity (\cite{dimova2002hyperviscous,rideau2018liposomes}), it means that the two-dimensional (2D) dissipation is larger than the three-dimensional one corresponding to large Boussinesq number $Bq$ as we will see further. Note that in polymersomes, the 2D dissipation is solely induced by the shear membrane viscosity (no impact of the dilational one) as the membrane is a 2D incompressible fluid. It is interesting to note that this is no longer the case for capsules.\\

Whether (micro-)capsules are artificial or (mimic) living cells, they are most often viscous droplets bounded by an elastic material membrane suspended in another viscous fluid that experiments under shear flow. The membrane of soft core-layer droplets such as capsules (cells) is made mainly of an assembly of polymers, surfactants, biopolymers and proteins (plasmalemma plus cortex in the last case) that interact strongly by cross-linking (\cite{chen2023robust,chu2011comparison,de2014mechanical}) or weakly by complexation (\cite{xie2017interfacial,rehage2002two,chachanidze2022structural,dubreuil2003elastic}). In the regime of small deformations, these examples lead to (micro-)capsules with mainly an elastic response with in some cases an additional viscous response. In the regime of large deformations, these examples of literature show two different behaviors, one dominated by an elastic response up to the breakup (\cite{chang1993experimental,chang1993experimentalb,de2015stretching,chu2011comparison}), and another exhibiting viscoplasticity (\cite{xie2017interfacial}). Here, we have focused on capsules which can withstand moderate and large deformation befor breaup and during it.

In section \ref{section:Generalities}, the basic equations (velocity continuity, mechanical equilibrium, Stokes model) driving fluid-structure interactions in the limit of low Reynolds number are recalled. Three constitutive laws are introduced: a strain-softening model (the neoHookean law), a strain-hardening model (the Skalak law) and the generalized Hooke model (\cite{pozrikidis2003modeling}). The salient dimensionless numbers, the elastic capillary number $Ca$ and the shear and dilational Boussinesq numbers $Bq_{s,d}$ leading to the two Weissenberg numbers $\beta_{s,d}$. In section \ref{section:Analytic}, the equation of the shape and the Lamb solutions of Stokes equations are recalled. In section \ref{section:ResultsCaSmall}, the shape and the dynamics are studied in the limit of small elastic capillary number associated to a finite effect of both membrane viscosities. In section \ref{section:PurelyViscous}, it is interesting to investigate the case without elasticity corresponding, in fact, to the limit $Ca \rightarrow \infty$. In section \ref{section:BqHigh}, the case with high membrane viscosities and finite capillary number is studied. In each limit, the transient dynamics is studied and the stationary state if exists is characterized. In section \ref{section:Discussion}, the results are summarized. We also highlight the importance of a potential difference between shear and dilational membrane viscosities. This has importance for any polymer layer of microcapsules but also for any biological cells.

\section{Problem statement and numerical procedure}
\label{section:Generalities}
\subsection{Problem statement}
\label{subsection:Problemstatement}
Throughout this section, we will introduce the bare physics involved here. Let a capsule, that is an initially spherical object with a thin elastic layer on its surface that obeys Hooke law. This object will contain a fluid of viscosity $\eta^*$, velocity $\bm{v}^*$ and pressure $p^*$ and be immersed in another fluid of viscosity $\eta$, velocity $\bm{v}$ and pressure $p$. Both fluids are Newtonian. 

Let $\bm{x}$ be a position vector of a point on the capsule in a cartesian basis where the origin is at the center of the capsule. Let $\bm{v}^\infty$ the velocity far from the capsule, that is, the undisturbed linear flow be
\begin{equation}
    \bm{v}^\infty = \dot\epsilon\left(\bm{E}+\bm{\Omega}\right)\cdot \bm{x}\ \text{for}\ |\bm{x}|\rightarrow\infty
    \label{eq:Vinfty}
\end{equation}
where $\dot\epsilon$ is the shear rate. $\bm{E}$ and $\bm{\Omega}$ are the decomposition into respectively symmetrical and anti symmetrical parts of the velocity gradient. For instance, in the case of shear flow one would have
\begin{equation}
    \bm{E}=\begin{pmatrix}
        0 & 1/2 & 0\\
        1/2 & 0 & 0\\
        0 & 0 & 0
    \end{pmatrix},\quad \bm{\Omega}=\begin{pmatrix}
        0 & 1/2 & 0\\
        -1/2 & 0 & 0\\
        0 & 0 & 0
    \end{pmatrix}
    \label{eq:matrixEOmega}
\end{equation}

The experiments in literature often consider a small capsule under hydrodynamic constraints in the limit of low Reynolds regime allowing to neglect any inertial contribution. Thus, the Stokes equations are relevant to solve the issue:
\begin{equation}
    \bm{\nabla}\cdot\bm{\Pi} = \bm{0}\qquad \bm{\nabla}\cdot\bm{v} = 0
\end{equation}
\begin{equation}
    \bm{\nabla}\cdot\bm{\Pi}^* = \bm{0}\qquad \bm{\nabla}\cdot\bm{v}^* = 0
\end{equation}
with the definition of the stress tensors $\bm{\Pi}$:
\begin{equation}
    \bm{\Pi} = -p\bm{I}+\eta\left(\bm{\nabla}\bm{v}+\bm{\nabla}\bm{v}^T\right)
\end{equation}
\begin{equation}
    \bm{\Pi}^* = -p^*\bm{I}+\eta^*\left(\bm{\nabla}\bm{v}^*+\bm{\nabla}\bm{v}^{*T}\right)
\end{equation}
The previous equations in the bulk are associated with boundary conditions. The first is the flow far from the capsule already given by equations \ref{eq:Vinfty} and \ref{eq:matrixEOmega} that leads to:
\begin{equation}
    \bm{v^\infty}\, =\, \dot\epsilon\, y\, \bm{e_x}
\end{equation}
where $\bm{e_x}$ is the unit vector aligned with the direction of the flow. $(x,y)$ is the shear plane while the vorticity is along the z axis in a Cartesian coordinate system. The other boundary equations apply at the membrane that bounds the inner fluid. The second is the continuity of fluid velocities: 
\begin{equation}
    \bm{v}+\bm{v}^\infty=\bm{v}^*\label{eq:velocity_continuity}
\end{equation}
where $\bm{v}$ is the variation in the fluid velocity due to the presence of the capsule.
\begin{equation}
    \left(\bm{\Pi}+\bm{\Pi}^\infty-\bm{\Pi}^*\right)\cdot\bm{n} = \bm{f}\label{eq:eq_force}
\end{equation}
where $\bm{n}$ is the normal vector on the capsule surface and pointing outwards. $\bm{f}$ is the sum of forces involved in our analysis and will be determined in the next section. The stress tensor at infinity $\bm{\Pi^\infty}$ is written as follows:
\begin{equation}
    \bm{\Pi}^\infty = -p^\infty\bm{I}+\eta\left(\bm{\nabla}\bm{v}^\infty+\bm{\nabla}\bm{v}^{\infty T}\right)
\end{equation}
where $p^\infty$ is an arbitrary constant. The last equation to introduce defines the evolution of the capsule shape over time:
\begin{equation}
    \frac{\partial r}{\partial t} = \left(\bm{v}+\bm{v}^\infty\right)\cdot\bm{n} 
    \label{eq:evolution}
\end{equation}

\subsection{Elastic force}
\label{sebsection:elasticforce}
The capsule membrane has the property similar to a spring meaning after its elongation, a comparison between the reference state of a point placed on the membrane and its final position is enforced. Let $\bm{X}$ be the reference position vector while $\bm{x}$ as previously said is the position vector; one can define the surface displacement gradient
\begin{equation}
    \bm{A} = \left(\bm{I}-\bm{n}\otimes \bm{n}^T\right)\cdot\frac{\bm{x}}{\bm{X}}\cdot\left(\bm{I}-\frac{\bm{X}}{|\bm{X}|}\otimes\frac{\bm{X}}{|\bm{X}|}\right)
\end{equation}
Defining the Green-Lagrange deformation tensor
\begin{equation}
    e = \frac{1}{2}\left[\bm{A}^T\cdot\bm{A}-\left(\bm{I}-\frac{\bm{X}}{|\bm{X}|}\otimes\frac{\bm{X}}{|\bm{X}|}\right)\right]
\end{equation}
one obtains the strain invariants and the Jacobian
\begin{equation}
    I_1 = 2\text{tr}\left(e\right)\qquad I_2 = J^2-1\qquad J = \sqrt{\det\left(\bm{A}^T\bm{A}+\frac{\bm{X}}{|\bm{X}|}\otimes\frac{\bm{X}}{|\bm{X}|}\right)}
\end{equation}
Based on the works from \cite{barthes2002effect}, the Cauchy tensor T is
\begin{equation}
    \bm{T} = \frac{2}{J}\left\{\frac{\partial w}{\partial I_1}\bm{A}\cdot\bm{A}^T+\frac{\partial w}{\partial I_2}J^2\left(\bm{I}-\bm{n}\otimes\bm{n}^T\right)\right\}
\end{equation}
where $w$ is the strain energy function and is dependent on the constitutive law used. in this framework, 3 laws will be used. Skalak law
\begin{equation}
    w^{Sk} = \frac{G^{Sk}_s}{4}\left(I_1^2+2I_1-2I_2+CI_2^2\right)
\end{equation}
is mostly used to described the strain exerted on red blood cells with $G_s^{Sk}$ the surface shear modulus and $C$ the Skalak parameter. Hooke law is the most common law as it's the same used for springs that is
\begin{equation}
    w^{H} = \frac{G^{H}_s}{4}\left(2I_1-2I_2+\frac{1}{1-\nu}I_1^2\right)
\end{equation}
for $G_s^H$ the surface shear modulus and $\nu$ the surface Poisson ratio. The last law studied here is the neo-hookean law that is a particular case of Kelvin-Voigt model written
\begin{equation}
    w^{NH} = \frac{G^{NH}_s}{2}\left(I_1-1+\frac{1}{I_2+1}\right)
\end{equation}
$G_s^{NH}$ is the surface shear elastic modulus.
The elastic obtained is then deduced by the surface divergence of the tensor $\bm{T}$ written
\begin{equation}
    \bm{f}_{el} = \bm{P}\cdot\bm{\nabla}\cdot\bm{T}
\end{equation}
where $\bm{P} = \bm{I}-\bm{n}\otimes\bm{n}^T$ is the projection tensor onto the surface. In the following, the surface shear elastic modulus will be denoted by $G_s$, regardless of the constitutive law used.
\subsection{Boussinesq-Scriven law}
\label{subsection:BSlaw}
In surface rheology (\cite{jaensson2021computational,sagis2011dynamic}), when the two-dimensional dissipation is expected to play a role, the most convenient way is to introduce the Boussinesq-Scriven model (\cite{boussinesq1913existence,scriven1960dynamics}). It is equivalent to a  compressible Newtonian fluid with a linear relation between the interfacial rate-of-deformation and the surface stress. This leads to the introduction of the dilational $\eta_d$ and shear $\eta_s$ surface viscosities. The Boussinesq-Scriven tensor $\bm{\Pi}^{BS}$ reads:
\begin{equation}
    \bm{\Pi}^{BS} = \eta_s\left[\bm{P}\cdot\nabla\bm{v}^*-\bm{P}\left(\nabla \bm{v}^*+\nabla \bm{v}^{*T}\right)\cdot\bm{P}\right]-\eta_d\bm{P}\cdot\nabla\bm{v}^*
    \label{eq:BS}
\end{equation}
The membrane force involved is then obtained by the expression
\begin{equation}
    \bm{f}^{BS} = \bm{f}_s^{BS} + \bm{f}_d^{BS} = \bm{P}\cdot\bm{\nabla}\cdot\bm{\Pi}^{BS}
\end{equation}

\subsection{Dimensionless quantities and relevant numbers}
\label{subsection:Dimensionless}
The elastic capillary number $Ca$ measures the magnitude of the viscous stress $\eta\dot\epsilon$ relative to the membrane elastic response $G_s/R$: 
\begin{equation}
    Ca\,=\,\frac{\eta \dot\epsilon R}{G_s}
\end{equation}
where $G_s$ depends on the constitutive law chosen. Before considering the two-dimensional rheology of the capsule membrane, the viscosity contrast is introduced to clarify the scope of this study:
\begin{equation}
    \lambda=\frac{\eta^*}{\eta}
\end{equation}
Indeed, in this paper, only the limit $\lambda << Bq$ is considered to highlight the contribution on shear and dilational membrane viscosities and for the sake of simplicity. This condition does not seem to appear to be mentioned in \cite{barthes1985role}. But from our point of view, this is an implicit assumption justified by the fact that the deformation under linear flow does not depend on the viscosity contrast up to the second order in the elastic capillary number $Ca$ (\cite{regazzi2026deformation}).

The two-dimensional and three-dimensional dissipations are compared evaluating the Boussinesq numbers for shear and dilational contributions:
\begin{equation}
    Bq_s\,=\,\frac{\eta_s}{\eta R}
    \label{eq:Bqs}
\end{equation}
\begin{equation}
    Bq_d\,=\,\frac{\eta_d}{\eta R}
    \label{eq:Bqd}
\end{equation}
In our framework, the two last dimensionless quantities are the ratio of the surface viscous stress $\eta_{s,d}\dot\epsilon$ to the elastic one $G_s/R$ or in other words, the membrane viscoelastic times multiplied by the shear rate or the ratio of the surface viscous stress to the elastic one:
\begin{equation}
    \beta_s\,=\,\frac{\eta_s}{G_s}\,\dot\epsilon
    \label{eq:betas}
\end{equation}
\begin{equation}
    \beta_d\,=\,\frac{\eta_d}{G_s}\,\dot\epsilon
    \label{eq:betad}
\end{equation}
with useful rewritings:
\begin{equation}
    \beta_s = Bq_s\, Ca\qquad \beta_d = Bq_d\, Ca
\end{equation}
are introduced. When the shear and dilational surface viscosities are equal, the notation becomes as follows:
\begin{equation}
    Bq = Bq_s = Bq_d
\end{equation}
\begin{equation}
    \beta = \beta_d = \beta_s = Bq\, Ca
\end{equation}
These numbers highlight the four times involved in the dynamics of a capsule under shear flow: the classic elastic characteristic time of a capsule $\eta R/G_s$, the surface dilational characteristic time of the membrane $\eta_d/G_s$, the surface shear one $\eta_s/G_s$ and finally, the characteristic time of the flow $1/\dot\epsilon$. To not choose arbitrarily a viscous relaxation time, the reference time used to dimensionless equations is $1/\dot\epsilon$. The reference length is the radius $R$, the stress reference is $\eta\dot\epsilon$.

When dealing with such equations as velocity continuity and mechanical equilibrium, working with non dimensional quantities is by-far easier. 
Mechanical equilibrium (equation \ref{eq:eq_force}) can be simplified to: 
\begin{equation}
    \left(\bm{\Pi}+\bm{\Pi}^\infty-\bm{\Pi}^*\right)\cdot\bm{n} = \frac{\bm{f}_{el}}{Ca}+\beta_s\frac{\bm{f}_s^{BS}}{Ca}+\beta_d\frac{\bm{f}^{BS}_d}{Ca}
    \label{eq:equimechAdim}
\end{equation}
where $\bm{\Pi}^* = -p^*\bm{I}+\lambda\left(\bm{\nabla}\bm{v}^*+\bm{\nabla}\bm{v}^{*T}\right)$ and $\lambda=\eta^*/\eta$, the viscosity contrast. $\bm{f}_s^{BS}$ and $\bm{f}_d^{BS}$ are the shear and dilational forces derived from the Boussinesq-Scriven tensor when $Bq_d=0$ and $Bq_s=0$ respectively (equation \ref{eq:BS}).\\

\section{Analytic solution}
\label{section:Analytic}
\setcounter{subsection}{-1}
\subsection{Notations}
We adopt the "truncated" Einstein convention, each repeated indices are summed over. We do not take into account the notion of covariant or contravariant indices positions. The superscript $*$ denotes the quantities expressed for the inner fluid.
\subsection{Position vector}
In this framework, only the small-deformation regime is concerned. The initially spherical capsule will be slightly deformed submitted to a shear flow. The slight change in shape is described as follows:
\begin{equation}
    r = 1+F = \sqrt{\bm{x}\cdot\bm{x}}
\end{equation}
with the condition $F<<1$. Based on the definition from \cite{barthes1981time}, the position vector of a point at the membrane surface is:
\begin{equation}
    \bm{x} = \bm{X} + 3 \bm{K}\cdot \bm{X}+3\left(\bm{F}-\bm{K}\right)\cdot\bm{X}\bm{X}\bm{X}/r^2
\end{equation}
where $K$ is a tensor describing the motion of a given point on the surface, it relates the initial and final position vectors. Naturally, $K<<1$ to satisfy the small deformation hypothesis.
This definition leads to the usual radius also written for a regular droplet where only the tensor $F$ is involved. Finally, the normal vector can be defined as
\begin{equation}
    \textbf{n}=\frac{\bm{\nabla} r-\bm{\nabla} F}{\vert\bm{\nabla} r-\bm{\nabla} F\vert}
\end{equation}

\subsection{Velocity fields}
\label{subsection:Lamb}
According to \cite{lamb1924hydrodynamics} Stokes equations have solutions involving solid harmonics and are written
\begin{equation}
    \bm{v} = \sum_{n=-\infty}^\infty \left[\bm{\nabla} \times \left(\bm{x} \chi^n\right)+\bm{\nabla} \phi^n + \frac{n+3}{2\left(n+1\right)\left(2n+3\right)}\bm{x}^2\bm{\nabla} p^n - \frac{n}{\left(n+1\right)\left(2n+3\right)}\bm{x}p^n\right]
\end{equation}
\begin{equation}
    \bm{v}^* = \sum_{n=-\infty}^\infty \left[\bm{\nabla} \times \left(\bm{x} \chi^{n,*}\right)+\bm{\nabla} \phi^{n,*} + \frac{n+3}{2\left(n+1\right)\left(2n+3\right)}\bm{x}^2\bm{\nabla} p^{n,*} - \frac{n}{\left(n+1\right)\left(2n+3\right)}\bm{x}p^{n,*}\right]
\end{equation}
$\chi$ is the rotating part of the fluid and $\phi$ the non-rotating part. $n$ is the order of solid harmonics. To know which harmonics is involved, a comparison with the far-field velocity and knowing that at far-field, the external velocity fades out, it becomes clear that harmonics of order $2$ remains.
Defining a tensor $\bm{T}$ for the pressure and $\bm{S}$ for the anti-rotating parts, the velocity fields are written locally
\begin{equation}
    v_i = 6 S_{li}X_lr^{-5}-15S_{lm}X_lX_mX_ir^{-7}+\frac{3}{2}T_{lm}X_lX_mX_ir^{-5}
\end{equation}
\begin{equation}
    v_i^* = 6 S_{li}^*X_l+\frac{5}{7}T_{li}^*X_lr^{2}-\frac{2}{7}T_{lm}^*X_lX_mX_i
\end{equation}
\subsection{Equations integration}
Since the capsule reach a steady state over time it does not deform furthermore. This leaves only the membrane rotating along time. The rotation with the lack of deformation allows to define Jaumann derivative that is a particular case of the Lie derivative
\begin{equation}
    \mathring{\bm{X}} = \dot{\bm{X}} - \bm{\omega}\cdot\bm{X} + \bm{X}\cdot\bm{\omega}
\end{equation}
where $\omega$ is antisymmetric and will be determined later.
To determine the velocity and stress fields, \cite{frankel1968motion} used a combinatoric method that is multiply by $\bm{x}$ and again with $\bm{x}\bm{x}\bm{x}$. The integration over a sphere volume $d\Omega$ gives the following results
\begin{equation}
    \int x_ix_jd\Omega = \frac{4\pi}{3}\delta_{ij}
\end{equation}
\begin{equation}
    \int x_ix_jx_lx_md\Omega = \frac{4\pi}{15}\left(\delta_{ij}\delta_{lm}+\delta_{il}\delta_{jm}+\delta_{im}\delta_{jl}\right)
\end{equation}
Applying this method for the deformation at order 0 (undeformed capsule) leads to
\begin{equation}
    \bm{\omega} = \bm{\Omega}+o(n)
\end{equation}
which means that $\bm{\omega}$ is an angular velocity. At first order, the evolution equation gives
\begin{equation}
    \frac{3}{15}T_{ij}+\frac{1}{3}E_{ij} = \mathring{K}_{ij}+\frac{6}{15}\left(\mathring{F}_{ij}-\mathring{K}_{ij}\right)
\end{equation}
\begin{equation}
    -9S_{ij}+\frac{3}{2}T_{ij}+E_{ij} = 3\mathring{K}_{ij}+3\left(\mathring{F}_{ij}-\mathring{K}_{ij}\right)
\end{equation}
Where the quantities from the velocity field are determined earlier and depends on the constitutive law. One can see that Jaumann derivative naturally comes in the expressions due to the vorticity terms in Lamb equations for the velocity fields. In local coordinates it is for instance
\begin{equation}
    \mathring{K}_{ij} = \dot{K}_{ij} + \frac{\omega_s}{2}\left( \varepsilon_{pjs}K_{ip}+\varepsilon_{pis}K_{jp} \right)
\end{equation}
Solving both equations can be tedious at first glance. By applying the method from \cite{barthes1985role} one defines the matrices
\begin{equation}
    \bm{\dot{M}} = \begin{pmatrix}
    \dot{\bm{F}}\\
    \dot{\bm{K}}
    \end{pmatrix},\qquad \bm{M} = \begin{pmatrix}
    \bm{F}\\
    \bm{K}
    \end{pmatrix}
\end{equation}
the following equation can the be written
\begin{equation}
    \bm{\dot{M}} = \bm{A} \bm{M} + \bm{B}\label{eq:matrix}
\end{equation}
It is clear that $\bm{B}$ is the matrix containing the solution at time $t\rightarrow\infty$. Based on the system of differential equations it creates, the solutions are quite straightforward.

\section{Results in the limit of small capillary number: $Ca<<1$ and $Bq_{s,d}=O(1)$}
\label{section:ResultsCaSmall}
Each subsection contains expressions for the tensors $\bm{F}$ and $\bm{K}$ rearranged to fit \cite{barthes1985role} expressions which were determined in the case of Skalak model, $C=1$ and equal Boussinesq numbers $Bq_s=Bq_d$. More generally, the results are also performed in the cases of neo-Hokean and Hooke models with different Boussinesq numbers. The temporal evolution of deformation and angle deviation (transitory to stationary or oscillating) and their final stationary values are identified the final state  and its properties. They come from solving the system of equations \ref{eq:velocity_continuity}, \ref{eq:eq_force} and \ref{eq:evolution}.\\
The tensors $\bm{K}$ and $\bm{F}$ are symmetric without trace that leads to useful relations:
\begin{equation}
    \bm{K}_{21}\,=\,\bm{K}_{12} \qquad \bm{K}_{22}\,=\,-\bm{K}_{11}
\end{equation}
\begin{equation}
    \bm{F}_{21}\,=\,\bm{F}_{12} \qquad \bm{F}_{22}\,=\,-\bm{F}_{11}
\end{equation}

The choice has been made to calculate independently each constitutive law case in this limit to allow any reader to easily work out again the temporal evolution of the shape, thanks to the expressions $F_{11}, F_{12}, K_{11}$ and $K_{12}$ for each constitutive law. \\
Here, we consider a small elastic capillary number and finite Boussinesq numbers: $Ca<<1$ and $Bq_{s,d}=O(1)$. In this limit, we will see that the shape evolves to a stationary ellipsoid with a fixed inclination compared to the direction (x-axis) of the shear flow. Both depend on the constitutive law. Finally, the membrane has a tank-treading motion and there is no transient oscillation.

\subsection{Skalak law}
\label{section:SkalakCaSmall}
\subsubsection{Temporal variation of the deformation: Skalak law}
Using Skalak law, after applying the initial condition, $r=1$ at $t=0$, the deformation $D_{12}$ in the plane $(x,y)$ (or (1,2)) can be simplified to:

\begin{equation}
    \begin{split}
        D_{12}^{Sk}(t)&=\frac{5}{8}Ca\biggl[\frac{16+12C(4+3C)+(3\beta_d+\beta_s)^2}{((1+2C)^2+\beta_d^2)(1+\beta_s^2)}+\frac{e^{-\frac{2t}{\tau_1^{Sk}}}}{(1+2C)^2+\beta_d^2}+\frac{9e^{-\frac{2t}{\tau_2^{Sk}}}}{1+\beta_s^2}\\
        &+6e^{-(\frac{1}{\tau_1^{Sk}}+\frac{1}{\tau_2^{Sk}})t}\frac{1+2C+\beta_d\beta_s}{((1+2C)^2+\beta_d^2)(1+\beta_s^2)}\\
        &+e^{-\frac{t}{\tau_2^{Sk}}}\frac{-6(4+2C(7+6C)+\beta_d(3\beta_d+\beta_s))\cos t+6(-\beta_d+\beta_s+2C\beta_s)\sin t}{((1+2C)^2+\beta_d^2)(1+\beta_s^2)}\\
        &+e^{-t/\tau_1^{Sk}}\frac{-2(4+6C+3\beta_d\beta_s+\beta_s^2)\cos t+6(\beta_d-(1+2C)\beta_s)\sin t}{((1+2C)^2+\beta_d^2)(1+\beta_s^2)}\biggr]^{1/2}
    \end{split}
\end{equation}
The deformation increases, resulting in a final shape that is an ellipsoid with a fixed orientation angle with the direction of the flow close to $\pi/4$ (see further). The variation from the initial spherical shape to the final one is driving by two characteristic times $\tau_{1,2}^{Sk}$ that depend on the Skalak law:\\
\begin{equation}
    \tau_1^{Sk}=\frac{\beta_d}{1+2C}
\end{equation}
\begin{equation}
    \tau_2^{Sk}=\beta_s
\end{equation}
These dimensionless times can be rewritten $T_{1,2}^{Sk}=\tau_{1,2}^{Sk}/\dot\epsilon$ in physical unities with the characteristic time $\eta R/G_s^{Sk}$ and Boussinesq numbers:\\
\begin{equation}
    T_1^{Sk}=\frac{\eta R}{G_s^{Sk}}\,\,\frac{Bq_d}{1+2C}\,=\,\frac{\eta_d}{G_s^{Sk}(1+2C)}\,=\,\frac{\eta_s}{K^{Sk}}
    \label{eq:TSkDilatCaSmall}
\end{equation}
where $K^{Sk}$ is nothing other than the surface dilation modulus (equations 1.3.10 and 1.3.11 of the chapter 1 of \cite{pozrikidis2003modeling}).
\begin{equation}
    T_2^{Sk}=\frac{\eta R}{G_s^{Sk}}\,\,Bq_s\,=\,\frac{\eta_s}{G_s^{Sk}}
    \label{eq:TSkshearCaSmall}
\end{equation}
These times do not depend on the bulk viscosity and on the radius. We recover the times discussed to introduced to define $\beta_s$ (equation \ref{eq:betas}) and $\beta_d$ (equation \ref{eq:betad}).\\

\subsubsection{Stationary state: Skalak law}
The final deformation, semi-axis lengths and orientation read:
\begin{equation}
   D_{12}^{Sk} = \frac{5}{8}\,Ca\,\sqrt{\frac{16+12C(4+3C)+(3\beta_d+\beta_s)^2}{\left((1+2C)^2+\beta_d^2\right)\left(1+\beta_s^2\right)}}
\end{equation}
\begin{equation}
    \frac{L^{Sk}}{R} = 1+\frac{5}{8}\,Ca\,\sqrt{\frac{16+12C(4+3C)+(3\beta_d+\beta_s)^2}{\left((1+2C)^2+\beta_d^2\right)\left(1+\beta_s^2\right)}}
\end{equation}
\begin{equation}
    \frac{S^{Sk}}{R} = 1-\frac{5}{8}\,Ca\,\sqrt{\frac{16+12C(4+3C)+(3\beta_d+\beta_s)^2}{\left((1+2C)^2+\beta_d^2\right)\left(1+\beta_s^2\right)}}
\end{equation}
\begin{equation}
    \frac{W^{Sk}}{R} = 1
\end{equation}
The deviation angle is also determined
\begin{equation}
\begin{split}
    \phi^{Sk} &= \frac{1}{2}\arctan \frac{2(1+2C)(2+3C)+3\beta_d^2+(1+2C)\beta_s^2}{3\left(1+2C\right)^2\beta_s+\beta_d+\beta_d\beta_s(3\beta_d+\beta_s)}\\
    &=\frac{\pi}{4}-\frac{1}{2}\arctan \frac{3\left(1+2C\right)^2\beta_s+\beta_d+\beta_d\beta_s(3\beta_d+\beta_s}{2(1+2C)(2+3C)+3\beta_d^2+(1+2C)\beta_s^2}
    \end{split}
\end{equation}\\
For $C=1$ and $\beta_d=\beta_s$, the stationary results from the seminal work of \cite{barthes1985role} are recovered. There is simply a difference of a factor $3$ coming from the definition of $G_s$.\\


\subsection{Hooke law}
\subsubsection{Temporal variation of the deformation: Hooke model}
As in the case of the Skalak law, the dynamics and the stationary state are similar. Using Hooke model, after applying the initial condition, $r=1$ at $t=0$, the deformation $D_{12}$ in the plane $(x,y)$ (or (1,2)) can be simplified to:
\begin{equation}
    \begin{split}
        D_ {12}^{H}(t) &= \frac{5}{8}Ca\biggl[\frac{16+(3\beta_d+\beta_s)^2-2\nu(-8+(3\beta_d+\beta_s)^2)+\nu^2(4+(3\beta_s+\beta_s)^2)}{((1+\nu)^2+(1-\nu)^2\beta_d^2)(1+\beta_s^2)}\\
        &+\frac{(-1+\nu)^2e^{-\frac{2t}{\tau_1^H}}}{(1+\nu)^2+(-1+\nu)^2\beta_d^2}+\frac{9e^{-\frac{2t}{\tau_2^H}}}{1+\beta_s^2}+6e^{-(\frac{1}{\tau_1^H}+\frac{1}{\tau_2^H}{})t}\frac{16\nu^2+(-1+\nu)^2\beta_d\beta_s}{((1+\nu)^2+(-1+\nu)^2\beta_d^2)(1+\beta_s^2)}\\
        &+e^{-\frac{t}{\tau_1^H}}\frac{2(-1+\nu)((4+3\beta_d\beta_s+\beta_s^2-\nu(-2+3\beta_d\beta_s+\beta_s^2))\cos t + 3((-1+\nu)\beta_d+(1+\nu)\beta_s)\sin t}{((1+\nu)^2+(-1+\nu)^2\beta_d^2)(1+\beta_s^2)}\\
        &+e^{-\frac{t}{\tau_2^H}}\frac{-6(4+2\nu(3+\nu)+3\beta_d^2+3(-2+\nu)\nu\beta_d^2+(-1+\nu)\beta_d\beta_s)\cos t}{((1+\nu)^2+(-1+\nu)^2\beta_d^2)(1+\beta_s^2)}\\
        &+e^{-\frac{t}{\tau_2^H}}\frac{-6(-1+\nu)((-1+\nu)\beta_d+(1+\nu)\beta_s)\sin t}{((1+\nu)^2+(-1+\nu)^2\beta_d^2)(1+\beta_s^2)}\biggr]^{1/2}
    \end{split}
\end{equation}
where the dimensionless characteristic times $\tau_{1,2}^{H}$ are:
\begin{equation}
    \tau_1^{H}=\frac{1-\nu}{1+\nu}\beta_d
\end{equation}
\begin{equation}
    \tau_2^{H}=\beta_s
\end{equation}
The two corresponding physical times $T_{1,2}^{H}=\tau_{1,2}^{H}/\dot\epsilon$ are:\\
\begin{equation}
    T_1^{H}=\frac{\eta R}{G_s^{Sk}}\,\,\frac{1-\nu}{1+\nu}Bq_d\,=\,\frac{1-\nu}{1+\nu}\frac{\eta_d}{G_s^{H}}\,=\,\frac{\eta_d}{K^{H}}
    \label{eq:THDilatCaSmall}
\end{equation}
where $K^{H}$ is  the surface dilation modulus.
\begin{equation}
    T_2^{H}=\frac{\eta R}{G_s^{H}}\,\,Bq_s\,=\,\frac{\eta_s}{G_s^{H}}
    \label{eq:THShearCaSmall}
\end{equation}
\subsubsection{Stationary state: Hooke model}
and the stationary state is
\begin{equation}
    D_{12}^{H} = \frac{5}{8}\,Ca\,\sqrt{\frac{16+(3\beta_d+\beta_s)^2-2\nu(-8+(3\beta_d+\beta_s)^2)+\nu^2(4+(3\beta_d+\beta_s)^2)}{((1+\nu)^2+(-1+\nu)^2\beta_d^2)(1+\beta_s^2)}}
\end{equation}
\begin{equation}
    \frac{L^H}{R} = 1+\frac{5}{8}\,Ca\,\sqrt{\frac{16+(3\beta_d+\beta_s)^2-2\nu(-8+(3\beta_d+\beta_s)^2)+\nu^2(4+(3\beta_d+\beta_s)^2)}{((1+\nu)^2+(-1+\nu)^2\beta_d^2)(1+\beta_s^2)}}
\end{equation}
\begin{equation}
    \frac{S^H}{R} = 1-\frac{5}{8}\,Ca\,\sqrt{\frac{16+(3\beta_d+\beta_s)^2-2\nu(-8+(3\beta_d+\beta_s)^2)+\nu^2(4+(3\beta_d+\beta_s)^2)}{((1+\nu)^2+(-1+\nu)^2\beta_d^2)(1+\beta_s^2)}}
\end{equation}
\begin{equation}
    \frac{W^H}{R} = 1
\end{equation}

\begin{equation}
\begin{split}
    \phi^{H} &= \frac{1}{2}\arctan \frac{2(1-\nu)(2-\nu)+3(1-\nu)^2\beta_d^2+(1-\nu^2)\beta_s^2}{3(1+\nu)^2\beta_s+(1-\nu)^2\beta_d+(1-\nu)^2\beta_s\beta_d(3\beta_d+\beta_s)}\\
    &=\frac{\pi}{4}-\frac{1}{2}\arctan\frac{3(1+\nu)^2\beta_s+(1-\nu)^2\beta_d+(1-\nu)^2\beta_s\beta_d(3\beta_d+\beta_s)}{2(1-\nu)(2-\nu)+3(1-\nu)^2\beta_d^2+(1-\nu^2)\beta_s^2}
    \end{split}
\end{equation}

\subsection{NeoHookean law}
\subsubsection{Temporal variation of deformation: neoHookean law}
As in the case of the Skalak law and Hooke model, the dynamics and the stationary state are similar. The main deformation in the plane $(1,2)$ (or $(x,y)$) reads:

\begin{equation}
    \begin{split}
        D_{12}^{NH}(t)&=\frac{5}{8}Ca\biggl[\frac{100+(3\beta_d+\beta_s)^2}{(9+\beta_d^2)(1+\beta_s^2)}+\frac{e^{-\frac{2t}{\tau_1^{NH}}}}{9+\beta_d^2}+\frac{9e^{-\frac{2t}{\tau_2^{NH}}}}{1+\beta_s^2}\\
        &+6e^{-t(\frac{1}{\tau_1^{NH}}+\frac{1}{\tau_2^{NH}})}\frac{3+\beta_d\beta_s}{(9+\beta_d^2)(1+\beta_s^2)}-6e^{-\frac{t}{\tau_2^{NH}}}\frac{((30+\beta_d(3\beta_d+\beta_s))\cos t+(\beta_d-3\beta_s)\sin t}{(9+\beta_d^2)(1+\beta_s^2)}\\
        &+e^{-\frac{t}{\tau_1^{NH}}}\frac{(-2(10+3\beta_d+\beta_s^2)\cos t+6(\beta_d-3\beta_s)\sin t}{(9+\beta_d^2)(1+\beta_s^2)}\biggr]
    \end{split}
\end{equation}
where the dimensionless characteristic times $\tau_{1,2}^{NH}$ are:
\begin{equation}
    \tau_1^{NH}=\frac{\beta_d}{3}
\end{equation}
\begin{equation}
    \tau_2^{NH}=\beta_s
\end{equation}
The two corresponding physical times $T_{1,2}^{Sk}=\tau_{1,2}^{Sk}/\dot\epsilon$ are:\\
\begin{equation}
    T_1^{NH}=\frac{\eta R}{G_s^{Sk}}\,\,\frac{Bq_d}{3}\,=\,\frac{\eta_d}{3 G_s^{NH}}\,=\,\frac{\eta_d}{K^{NH}}
    \label{eq:TNHDilatCaSmall}
\end{equation}
where $K^{NH}$ is  the surface dilation modulus .
\begin{equation}
    T_2^{NH}=\frac{\eta R}{G_s^{NH}}\,\,Bq_s\,=\,\frac{\eta_s}{G_s^{NH}}
    \label{eq:TNHShearCaSmall}
\end{equation}

\subsubsection{Stationary state: neoHookean law}
the stationary state
\begin{equation}
    \begin{split}
    D_{12}^{NH} = \frac{5}{8}\,Ca\,\sqrt{\frac{100+(3\beta_d+\beta_s)^2}{\left(1+\beta_s^2\right)\left(9+\beta_d^2\right)}}
    \end{split}
\end{equation}
with the lengths
\begin{equation}
    \frac{L^{NH}}{R} = 1+\frac{5}{8}\,Ca\,\sqrt{\frac{100+(3\beta_d+\beta_s)^2}{\left(1+\beta_s^2\right)\left(9+\beta_d^2\right)}}
\end{equation}
\begin{equation}
    \frac{S^{NH}}{R} = 1-\frac{5}{8}\,Ca\,\sqrt{\frac{100+(3\beta_d+\beta_s)^2}{\left(1+\beta_s^2\right)\left(9+\beta_d^2\right)}}
\end{equation}
\begin{equation}
    \frac{W^{NH}}{R} = 1
\end{equation}
while and the deviation angle
\begin{equation}
\begin{split}
    \phi^{NH} &=\frac{1}{2}\arctan \frac{3\left(10+\beta_d^2+\beta_s^2\right)}{27\beta_s+\beta_d+\beta_s\beta_d(3\beta_d+\beta_s)}\\
    &=\frac{\pi}{4}-\frac{1}{2}\arctan \frac{27\beta_s+\beta_d+\beta_s\beta_d(3\beta_d+\beta_s)}{3(10+\beta_d^2+\beta_s^2)}
        \end{split}
\end{equation}

\section{Results in the case of purely viscous membrane}
\label{section:PurelyViscous}
Before dealing with the case of large Boussinesq numbers and finite capillary number, is is interesting to consider the case without elasticity: $\bm{f}_{el} =\bm{0}$ in equation \ref{eq:equimechAdim}. It also corresponds to the limit when the capillary number grows to infinity while $\beta_{s,d}/Ca=Bq_{s,d}$ are constant.
Thus, there is no point in distinguishing between the constitutive laws as in previous section. Based on \cite{barthes1985role}, solving the evolution equation for capsule deformation leads to the following quantities
\begin{equation}
    \begin{pmatrix}
        F_{11}\\
        F_{12}\\
    \end{pmatrix} = \frac{5\left(3Bq_d + Bq_s\right)}{24 Bq_d Bq_s}\begin{pmatrix}
        1-\cos t\\
        \sin t
    \end{pmatrix}\qquad\quad
     \begin{pmatrix}
        K_{11}\\
        K_{12}\\
    \end{pmatrix} = \frac{5}{12Bq_s}\begin{pmatrix}
        1-\cos t\\
        \sin t
    \end{pmatrix}
\end{equation}
Whatever the constitutive law in consideration, the principal lengths of the equivalent ellipsoid are
\begin{equation}
    \frac{L(t)}{R} = 1+\frac{5\left(3Bq_d + Bq_s\right)}{4 Bq_d Bq_s}|\sin \frac{t}{2}|
\end{equation}
\begin{equation}
    \frac{S(t)}{R} = 1-\frac{5\left(3Bq_d + Bq_s\right)}{4 Bq_d Bq_s}|\sin \frac{t}{2}|
\end{equation}
\begin{equation}
    \frac{W(t)}{R} = 1
\end{equation}
leading to the deformation
\begin{equation}
    D_{12}(t) = \frac{5\left(3Bq_d + Bq_s\right)}{4 Bq_d Bq_s}|\sin \frac{t}{2}|
\end{equation}
Asymptotic analysis shows that $D_{12}$ only depends on $Bq_d$ if $Bq_s$ is larger. The reciprocal is also valid, dependence on $Bq_s$ when $Bq_d$ growing faster. Figures \ref{fig:deformation_bqs_fixed} and \ref{fig:deformation_bqd_fixed} display the deformation of the capsule that oscillates along time according to fixed values for the viscosities. One can see again that $Bq_s$ contributes more in the deformation than $Bq_d$ according to the factor 3.\\

The angle of inclination oscillates at a frequency proportional to the shear rate and does not depend on the Boussinesq numbers.
\begin{equation}
    \theta = \frac{\pi}{4}-\frac{t}{4},\qquad t\in\left[0,2\pi\right]
\end{equation}

In the case of equal Boussinesq numbers $Bq_s=Bq_d$, these results are in agreement with \cite{barthes1985role} in the case $m=5$ (notation of the cited paper, equation $7.2$) for an elastomer elasticity. 

\section{Results in the case of high membrane viscosity: $Bq>>1$ and $Ca=O(1)$}
\label{section:BqHigh}
The capsule has now a high viscous membrane with a finite elasticity. The elastic time $\eta R /G_s$ is finite that can relax the permanent oscillations observed in the previous section \ref{section:PurelyViscous}. A new behavior is expected. We will demonstrate that this dynamics corresponds to a damping oscillation to a final tank-treading motion different from the one in the limit $Ca<<1$ (section \ref{section:ResultsCaSmall}).\\
\subsection{Temporal variation of the deformation}
The solutions are found using the same method as before. The deformation $D_{12}^{Law}$ in the plane $(x,y)$ (or $(1,2)$) can be written in a general way whatever the constitutive law as:
\begin{equation}
    \begin{split}
        D_{12}^{Law}(t) &= \frac{5(3Bq_d+Bq_s)}{8Bq_dBq_s}\Biggl(1
        -2e^{-t/\tau^{Law}}\cos(t)\left[\cosh\left(\chi^{Law}t\right)-2\sinh\left(\chi^{Law}t\right)\right]\\
        & +e^{-t/2\tau^{Law}}\left[\cosh\left(\chi^{Law}t\right)-2\sinh\left(\chi^{Law}t\right)\right]^2\Biggr)^{1/2}
    \end{split}
    \label{eq:D12BqHigh}
\end{equation}
where $\tau^{Law}$ and $\chi^{Law}$ are two dimensionless quantities. Both depend on the constitutive law.\\
This dynamics is a damped oscillation that leads to a tank-treading motion. The transient is completely different from the case $Ca<<1$ in section \ref{section:ResultsCaSmall}. Initially, the capsule is spherical. The capsule shape deforms into an ellipsoidal shape that oscillates at a frequency equal to the shear rate; its orientation with respect to the direction of the flow also oscillates. The deformation and the orientation angle relax to a stationary value over the dimensionless characteristic times $\tau^{Law}$ and $2\,\tau^{Law}$. The final stationary deformation does not depend on the constitutive law while the final orientation angle does. 

\subsubsection{Damping characteristic time with Skalak law}
The dimensionless characteristic times $\tau_{1,2}^{Sk}$ of the damping can be written in terms of physical times $T_{1,2}^{Sk}=\tau_{1,2}^{Sk}/\dot\epsilon$ proportional to the elastic time $\eta R/G_s^{Sk}$:
\begin{equation}
    T_1^{Sk} = 2T_2^{Sk} = \frac{\eta R}{G_s^{Sk}}\,\frac{Bq_d Bq_s}{Bq_d+Bq_s(1+2C)}
\end{equation}
To close the solution, the dimensionless parameter $\chi^{Sk}$ reads:
\begin{equation}
    \chi^{Sk}=\frac{Bq_d-Bq_s(1+2C)}{2Bq_dBq_sCa}
\end{equation}

\subsubsection{Damping characteristic time with neoHookean law}
The dimensionless characteristic times $\tau_{1,2}^{NH}$ of the damping can be written in terms of physical times $T_{1,2}^{NH}=\tau_{1,2}^{NH}/\dot\epsilon$ proportional to the elastic time $\eta R/G_s^{NH}$:
\begin{equation}
    T_1^{NH} = 2T_2^{NH} = \frac{\eta R}{G_s^{NH}}\,\frac{Bq_d Bq_s}{Bq_d+3Bq_s}
\end{equation}
To close the solution, the dimensionless parameter $\chi^{NH}$ reads:
\begin{equation}
    \chi^{NH}=\frac{Bq_d-3Bq_s}{2Bq_dBq_s Ca}
\end{equation}

\subsubsection{Damping characteristic time with Hooke model}
The dimensionless characteristic times $\tau_{1,2}^{H}$ of the damping can be written in terms of physical times $T_{1,2}^{H}=\tau_{1,2}^{H}/\dot\epsilon$ proportional to the elastic time $\eta R/G_s^{H}$:
\begin{equation}
    T_1^{H} = 2T_2^{H} = \frac{\eta R}{G_s^{H}}\,\frac{Bq_d Bq_s (1-\nu)}{Bq_d(1-\nu)+Bq_s(1+\nu)}
\end{equation}
To close the solution, the dimensionless parameter $\chi^{H}$ reads:
\begin{equation}
    \chi^{H}=\frac{Bq_d(1-\nu)-Bq_s(1+\nu)}{2Bq_dBq_s (1-\nu) Ca}
\end{equation}

\subsection{Final stationary state deformations, lengths and orientation angle}
As shown by equation \ref{eq:D12BqHigh}, the deformation is constant after the damping time and depends only on the Boussinesq numbers. It is also the case of the lengths $(L,S,W)$ of the semi-axis of the ellipsoidal shape:
\begin{equation}
    D_{12} = \frac{5(3Bq_d+Bq_s)}{8Bq_dBq_s}
    \label{eq:D12BqHighStat}
\end{equation}
\begin{equation}
    \frac{L}{R} =1+\frac{5(3Bq_d+Bq_s)}{8Bq_dBq_s}
\end{equation}
\begin{equation}
    \frac{S}{R} =1-\frac{5(3Bq_d+Bq_s)}{8Bq_dBq_s}
\end{equation}
\begin{equation}
    \frac{W}{R} =1
\end{equation}
\begin{equation}
    D_{13} = -D_{23} = \frac{1}{2}D_{12}
\end{equation}
This last relation is solely satisfied at the first order of the inverse of Boussinesq number ($1/Bq<<1$).
The stationary orientation angle of the capsule with the direction of the flow depends on the constitutive law:
\begin{itemize}
    \item Skalak law
    \begin{equation}
    \phi_{TT}^{Sk}=\frac{3Bq_d-Bq_s(1+2C)}{4Bq_dBq_s Ca}
    \end{equation}
    \item NeoHookean law
    \begin{equation}
    \phi_{TT}^{NH}=\frac{3\left(Bq_d-Bq_s\right)}{4Bq_dBq_sCa}
    \end{equation}
    \item Hooke model
    \begin{equation}
    \phi_{TT}^{H}=\frac{3Bq_d(1-\nu)-Bq_s(1+\nu)}{4Bq_dBq_s (1-\nu)Ca}
    \end{equation}
\end{itemize}

\section{Discussion}
\label{section:Discussion}
In the limit of evanescent membrane viscosities ($\beta_{s,d}\rightarrow 0$), we recover at the first order in the elastic capillary number ($Ca<<1$), the same expressions of deformations and lengths of the capsule  and its orientation angle equal to $\pi/4$. This asymptotic limit is satisfied whatever the constitutive law.\\

Three dynamics are identified depending on the values of the elastic capillary number $Ca$ and the Boussinesq numbers $Bq_{s,d}$:
\begin{itemize}
    \item Permanent oscillations in the case without elasticity (purely viscous membrane): this oscillatory dynamics at the frequency equal to the shear rate corresponds to the limit $Ca\rightarrow\infty$. More physically, if $T_{exp}$ is the time over which the capsule’s movement is analysed, this oscillation is observed if $T_{exp}<T_{1,2}^{Law}$ where $T_{1,2}^{Law}$ is the damping time that still exists for real membranes. Meanwhile, the amplitudes do not depend on the constitutive laws and only on the Boussinesq numbers, which means the membrane viscosities. At this first order of analysis in $1/Bq$, It is not possible to determine the two Boussinesq numbers by tracking how the lengths change over time. 
    \item Tank-treading motion beginning with damped oscillations at a frequency equal to the shear rate in the limit of high Boussinesq number: $Bq>>1$ and $Ca=O(1)$: this case is the same as the previous one, but with finite elasticity $G_s\neq0$, an essential parameter. The oscillating shape relaxes to an ellipsoidal one at a fixed orientation angle with the direction of the flow. The final deformation does not depend on the constitutive law and on the elastic capillary number while the orientation angle does. The variation in deformations or lengths is greater when the Boussinesq shear number $Bq_s$ varies than when the dilational Boussinesq number does. Meanwhile, the orientation angle can be negative in this limit of high membrane viscosity if the shear Boussinesq number is higher than the dilational one (neoHookean law and Hooke model) or if the membrane is strongly strain-hardening (large value of the parameter $C$ of the Skalak law). Note that such an unexpected behavior is also observed for a Skalak capsule when the viscosity contrast is very high (\cite{regazzi2026deformation}). Finally, if the constitutive law is known or can be assumed in a first approximation to interpret some experimental results, the monitoring of the deformation and the orientation angle in the stationary state allows to determine both Boussinesq numbers and so the two membrane viscosities.
    \item Tank-treading motion through the limit of small capillary number: $Ca<<1$ and $Bq=O(1)$: it appears smoothly without any damped oscillation and likely any overshoot contrary to the previous tank-treading motion in the limit of high membrane viscosities. The initial spherical shape deforms to an ellipsoidal one at a fixed orientation angle with the direction of the flow which gives also the sense of membrane rotation (axis of vorticity). The stationary deformations, lengths, orientation angle and the dynamics depend on the constitutive law. In particular, the deformations and lengths are proportional to the elastic capillary number contrary to the tank-treading motion in the limit of high viscous membrane ($Bq>>1$) for which the deformation does not depend on $Ca$. The correction of orientation angle to the value $\pi/4$ is proportional to $Ca$ in the limit $\beta=Bq\,Ca<<1$ while in the limit $Bq>>1$, the orientation angle is proportion to the inverse of the product $Bq\,Ca$ demonstrating once again the different physical origin of both tank-treading motions. By introducing the surface dilational elastic modulus $K^{Law}$, all characteristic times can be simply expressed as the ratio of the membrane's dilational (shear) viscosity to its dilational (shear) modulus of elasticity: equations \ref{eq:TSkDilatCaSmall}, \ref{eq:TSkshearCaSmall}, \ref{eq:THDilatCaSmall}, \ref{eq:THShearCaSmall}, \ref{eq:TNHDilatCaSmall} and \ref{eq:TNHShearCaSmall}. It would have been possible to introduce these quantities from the outset by considering Weissenberg numbers, which, from our perspective, are more $\dot\epsilon\,T^{Law}_{shear}$ and $\dot\epsilon\,T^{Law}_{dilational}$ than the numbers $\beta$ (equations \ref{eq:betas}, \ref{eq:betad}). The times are given by the previous equations cited. Finally, measuring the deformations (or lengths) and orientation angle should lead to the determination of the two surface viscosities, given that the capillary number can be varied to expand the range of parameters. 
\end{itemize}

    Interestingly, the expression of the deformation $D_{1,2}$ determined in the limit $Ca<<1$ and $Bq=O(1)$ provides the same expression of the deformation in the limit $Bq>>1$ when $Bq_s$ and $Bq_d$ tend to large values. Then, to discuss more, we can speculate that the equation in the limit $Ca<<1$ is valid for any value of the couple $(Bq_s,Bq_d)$. Indeed, there is no physical reason for the shear and dilational surface viscosities to have the same order of magnitude (\cite{zell2014surface}). Let us take the Hooke model for the constitutive law of the capsule membrane:
    \begin{itemize}
    \item $\beta_s>>1$ and $\beta_d=O(1)$ 
    \begin{equation}
    D_{12}^H \approx \frac{5}{8}\,Ca\,\sqrt{\frac{\left(1-\nu\right)^2}{\left(1+\nu\right)^2+\beta_d^2\left(1-\nu\right)^2}}
    \label{eq:D12BetasHigh}
    \end{equation}
    Consider again the physical characteristic time $T_{exp}$ of the experiment. If $T_{exp} << T_2^H=\eta_s/G_s^H=\beta_s/\dot\epsilon$, the deformation depends only on the dilational viscosity and the capillary number: equation \ref{eq:D12BetasHigh}. On the contrary, the equation \ref{eq:D12BqHighStat} has to be applied.
    \item $\beta_d>>1$ and $\beta_s=O(1)$
    \begin{equation}
    D_{12}^H \approx \frac{15}{8\left(1+\beta_s\right)}\,Ca
    \label{eq:D12BetadHigh}
    \end{equation}
    If $T_{exp} << T_1^H=\eta_d/K_s^H$, the deformation depends only on the shear viscosity and the capillary number: equation \ref{eq:D12BetadHigh}. In contrast, equation \ref{eq:D12BqHighStat} has to be applied.
    \end{itemize}
 
\section{Conclusion}
The shape and dynamics of soft core-layer microparticles depend on the rheological properties of the viscoelastic thin layer whether the membrane is fluid or solid. As this layer often results from a complex physico-chemical process (curing, interfacial polymerisation...), the knowledge and the characterisation of these viscoelastic thin films are poorly understood, in the linear regime of deformation and also in the one of large deformations which depends on the constitutive law (Hooke, strain-hardening or strain softening models). Nevertheless, these properties are fundamental to improving them and designing new, more efficient membranes. This remains a challenge.\\

In this study, the deformation of a microcapsule with a viscoelastic membrane is investigated under shear flow in the regime of weak deformation and in the limit of low Reynolds number. Both shear and dilational membrane viscosities are introduced. Contrary to a vesicle whose the membrane is a two-dimensional incompressible fluid, the contribution of the dilational viscosity has no reason to be neglected in the case of microcapsules. If the limit without elasticity is ignored, the microcapsule (initially spherical) has only a tank-treading motion with two different modes, one that depends on the capillary number $Ca$ in the limit of small $Ca$, another in the limit of high membrane viscosities does not. As shown in the discussion, this picture is more complex when the dilational and shear membrane viscosities are different. These results provide efficient tools to gain insight in the interfacial rheology of thin layers between two liquids.\\

Finally, as the seminal work of \cite{barthes1985role} was a solid basis to validate a numerical code of viscoelastic interfaces in the case of Skalak model with $C=1$ and $Bq_s=Bq_d$, the new results presented here allow to generalize to any constitutive law and to the essential case: $Bq_s\,\neq\,Bq_d$. This is particularly interesting when seeking to describe the behavior of biological cells or spheroids and beyond, or even systems with complex physicochemical properties.\\

\section{acknowledgment}
Paul Regazzi received
support from the french government under the France 2030 investment plan, as part of the
Initiative d’Excellence d’Aix-Marseille Université – AMIDEX and the CNRS through PhD grant. We also acknowledge for funding support from the
French Agence Nationale de la Recherche in the
framework of project SOCA, ANR-24-CE06-6245 and the National Centre for Space Studies, CNES. Centre de Calcul Intensif d’Aix-Marseille is acknowledged for granting access to its high performance computing resources.
\bibliographystyle{jfm}
\bibliography{biblio_these}

@article{boussinesq1913existence,
  title={Sur l’existence d’une viscosit{\'e} seperficielle, dans la mince couche de transition s{\'e}parant un liquide d’un autre fluide contigu},
  author={Boussinesq, MJ},
  journal={Ann. Chim. Phys.},
  volume={29},
  pages={349},
  year={1913}
}

@article{zell2014surface,
  title={Surface shear inviscidity of soluble surfactants},
  author={Zell, Zachary A and Nowbahar, Arash and Mansard, Vincent and Leal, L Gary and Deshmukh, Suraj S and Mecca, Jodi M and Tucker, Christopher J and Squires, Todd M},
  journal={Proceedings of the National Academy of Sciences},
  volume={111},
  number={10},
  pages={3677--3682},
  year={2014},
  publisher={National Academy of Sciences}
}

@article{jaensson2021computational,
  title={Computational interfacial rheology},
  author={Jaensson, Nick O and Anderson, Patrick D and Vermant, Jan},
  journal={Journal of Non-Newtonian Fluid Mechanics},
  volume={290},
  pages={104507},
  year={2021},
  publisher={Elsevier}
}

@article{sagis2011dynamic,
  title={Dynamic properties of interfaces in soft matter: Experiments and theory},
  author={Sagis, Leonard M},
  journal={Reviews of Modern Physics},
  volume={83},
  number={4},
  pages={1367},
  year={2011}
}

@article{scriven1960dynamics,
  title={Dynamics of a fluid interface equation of motion for Newtonian surface fluids},
  author={Scriven, Laurence E},
  journal={Chemical Engineering Science},
  volume={12},
  number={2},
  pages={98--108},
  year={1960},
  publisher={Elsevier}
}

@article{barthes1985role,
  title={Role of membrane viscosity in the orientation and deformation of a spherical capsule suspended in shear flow},
  author={Barthes-Biesel, Dominique and Sgaier, H},
  journal={Journal of Fluid Mechanics},
  volume={160},
  pages={119--135},
  year={1985},
  publisher={Cambridge University Press}
}

@article{chachanidze2022structural,
  title={Structural characterization of the interfacial self-assembly of chitosan with oppositely charged surfactant},
  author={Chachanidze, Revaz and Xie, Kaili and Massaad, Hanna and Roux, Denis and Leonetti, Marc and de Loubens, Cl{\'e}ment},
  journal={Journal of colloid and interface science},
  volume={616},
  pages={911--920},
  year={2022},
  publisher={Elsevier}
}

@book{pozrikidis2003modeling,
  title={Modeling and simulation of capsules and biological cells},
  author={Pozrikidis, Constantine},
  year={2003},
  publisher={CRC Press}
}

@article{chatkaew2009dynamics,
  title={Dynamics of vesicle unbinding under axisymmetric flow},
  author={Chatkaew, Sunita and Georgelin, Marc and Jaeger, Marc and Leonetti, Marc},
  journal={Physical review letters},
  volume={103},
  number={24},
  pages={248103},
  year={2009},
  publisher={APS}
}

@article{iyer2022non,
  title={Non-equilibrium shapes and dynamics of active vesicles},
  author={Iyer, Priyanka and Gompper, Gerhard and Fedosov, Dmitry A},
  journal={Soft matter},
  volume={18},
  number={36},
  pages={6868--6881},
  year={2022},
  publisher={The Royal Society of Chemistry}
}

@article{lipowsky2012spontaneous,
  title={Spontaneous tubulation of membranes and vesicles reveals membrane tension generated by spontaneous curvature},
  author={Lipowsky, Reinhard},
  journal={Faraday discussions},
  volume={161},
  pages={305--331},
  year={2012},
  publisher={The Royal Society of Chemistry}
}

@article{young2026soft,
  title={Soft-Lubrication Drainage and Rupture in Particle-Driven Vesicles},
  author={Young, Yuan-Nan and Quaife, Bryan and Nganguia, Herve and Pak, On Shun and Feng, Jie and Stone, Howard A},
  journal={Physical review letters},
  volume={136},
  number={20},
  pages={204001},
  year={2026},
  publisher={APS}
}

@article{boedec2013sedimentation,
  title={Sedimentation-induced tether on a settling vesicle},
  author={Boedec, Gwenn and Jaeger, Marc and Leonetti, Marc},
  journal={Physical Review E—Statistical, Nonlinear, and Soft Matter Physics},
  volume={88},
  number={1},
  pages={010702},
  year={2013},
  publisher={APS}
}

@article{shishkin2026metastability,
  title={Metastability of lipid vesicle shapes in uniaxial extensional flow},
  author={Shishkin, MA and Pikina, ES},
  journal={Physical Review E},
  volume={113},
  number={5},
  pages={055417},
  year={2026},
  publisher={APS}
}

@article{smith2004pulling,
  title={Pulling tethers from adhered vesicles},
  author={Smith, Ana-Sun{\v{c}}ana and Sackmann, Erich and Seifert, Udo},
  journal={Physical review letters},
  volume={92},
  number={20},
  pages={208101},
  year={2004},
  publisher={APS}
}

@article{vlahovska2007dynamics,
  title={Dynamics of a viscous vesicle in linear flows},
  author={Vlahovska, Petia M and Gracia, Ruben Serral},
  journal={Physical Review E—Statistical, Nonlinear, and Soft Matter Physics},
  volume={75},
  number={1},
  pages={016313},
  year={2007},
  publisher={APS}
}

@article{lyu2023swirling,
  title={Swirling of vesicles: Shapes and dynamics in Poiseuille flow as a model of RBC microcirculation},
  author={Lyu, Jinming and Chen, Paul G and Farutin, Alexander and Jaeger, Marc and Misbah, Chaouqi and Leonetti, Marc},
  journal={Physical Review Fluids},
  volume={8},
  number={2},
  pages={L021602},
  year={2023},
  publisher={APS}
}

@article{rideau2018liposomes,
  title={Liposomes and polymersomes: a comparative review towards cell mimicking},
  author={Rideau, Emeline and Dimova, Rumiana and Schwille, Petra and Wurm, Frederik R and Landfester, Katharina},
  journal={Chemical society reviews},
  volume={47},
  number={23},
  pages={8572--8610},
  year={2018},
  publisher={The Royal Society of Chemistry}
}

@article{zhang2017polymersomes,
  title={Polymersomes in nanomedicine-a review},
  author={Zhang, Xiao-ying and Zhang, Pei-ying},
  journal={Current Nanoscience},
  volume={13},
  number={2},
  pages={124--129},
  year={2017},
  publisher={Bentham Science Publishers}
}

@article{biben2011three,
  title={Three-dimensional vesicles under shear flow: Numerical study of dynamics and phase diagram},
  author={Biben, Thierry and Farutin, Alexander and Misbah, Chaouqi},
  journal={Physical Review E—Statistical, Nonlinear, and Soft Matter Physics},
  volume={83},
  number={3},
  pages={031921},
  year={2011},
  publisher={APS}
}

@article{agarwal2020stable,
  title={Stable shapes of three-dimensional vesicles in unconfined and confined Poiseuille flow},
  author={Agarwal, Dhwanit and Biros, George},
  journal={Physical Review Fluids},
  volume={5},
  number={1},
  pages={013603},
  year={2020},
  publisher={APS}
}

@article{kaoui2009red,
  title={Why do red blood cells have asymmetric shapes even in a symmetric flow?},
  author={Kaoui, Badr and Biros, George and Misbah, Chaouqi},
  journal={Physical review letters},
  volume={103},
  number={18},
  pages={188101},
  year={2009},
  publisher={APS}
}

@article{rangamani2013lipid,
  title={Lipid tubule growth by osmotic pressure},
  author={Rangamani, Padmini and Zhang, Di and Oster, George and Shen, Amy Q},
  journal={Journal of The Royal Society Interface},
  volume={10},
  number={88},
  pages={20130637},
  year={2013}
}

@article{waugh2009membrane,
  title={Membrane tethers},
  author={Waugh, Richard E},
  journal={Current Topics in Membranes},
  volume={64},
  pages={3--24},
  year={2009},
  publisher={Elsevier}
}

@article{seifert1997configurations,
  title={Configurations of fluid membranes and vesicles},
  author={Seifert, Udo},
  journal={Advances in physics},
  volume={46},
  number={1},
  pages={13--137},
  year={1997},
  publisher={Taylor \& Francis}
}

@article{dimova2002hyperviscous,
  title={Hyperviscous diblock copolymer vesicles},
  author={Dimova, R and Seifert, U and Pouligny, B and F{\"o}rster, S and D{\"o}bereiner, H-G},
  journal={The European physical journal E},
  volume={7},
  number={3},
  pages={241--250},
  year={2002},
  publisher={Springer}
}

@article{waugh1982surface,
  title={Surface viscosity measurements from large bilayer vesicle tether formation. II. Experiments},
  author={Waugh, RICHARD E},
  journal={Biophysical journal},
  volume={38},
  number={1},
  pages={29--37},
  year={1982},
  publisher={Elsevier}
}

@article{regazzi2026deformation,
  title={Theory of capsule deformation and orientation under shear flow: tank-treading, damped and permanent oscillations},
  author={Regazzi, Paul and Leonetti, Marc},
  journal={submitted to J. Fluid Mech.},
  year={2026}
}

@article{dionzou2016comparison,
  title={Comparison of methods for the fabrication and the characterization of polymer self-assemblies: what are the important parameters?},
  author={Dionzou, M and Mor{\`e}re, A and Roux, Cl{\'e}ment and Lonetti, Barbara and Marty, J-D and Mingotaud, Christophe and Joseph, Pierre and Goudoun{\`e}che, D and Payr{\'e}, B and L{\'e}onetti, M and others},
  journal={Soft matter},
  volume={12},
  number={7},
  pages={2166--2176},
  year={2016},
  publisher={Royal Society of Chemistry}
}

@article{de2015stretching,
  title={Stretching of capsules in an elongation flow, a route to constitutive law},
  author={De Loubens, Clement and Deschamps, Julien and Boedec, Gwenn and Leonetti, Marc},
  journal={Journal of Fluid Mechanics},
  volume={767},
  pages={R3},
  year={2015},
  publisher={Cambridge University Press}
}

@article{chu2011comparison,
  title={Comparison between measurements of elasticity and free amino group content of ovalbumin microcapsule membranes: discrimination of the cross-linking degree},
  author={Chu, TX and Salsac, A-V and Leclerc, E and Barth{\`e}s-Biesel, D and Wurtz, H and Edwards-L{\'e}vy, F},
  journal={Journal of colloid and interface science},
  volume={355},
  number={1},
  pages={81--88},
  year={2011},
  publisher={Elsevier}
}

@article{dubreuil2003elastic,
  title={Elastic properties of polyelectrolyte capsules studied by atomic-force microscopy and RICM},
  author={Dubreuil, Frederic and Elsner, Nils and Fery, Andreas},
  journal={The European Physical Journal E},
  volume={12},
  number={2},
  pages={215--221},
  year={2003},
  publisher={Springer}
}

@article{rehage2002two,
  title={From two-dimensional model networks to microcapsules},
  author={Rehage, Heinz and Husmann, Martin and Walter, Anja},
  journal={Rheologica acta},
  volume={41},
  number={4},
  pages={292--306},
  year={2002},
  publisher={Springer}
}

@article{chang1993experimental,
  title={Experimental studies of the deformation and breakup of a synthetic capsule in steady and unsteady simple shear flow},
  author={Chang, Keun-Shik and Olbricht, William Lee},
  journal={Journal of Fluid Mechanics},
  volume={250},
  pages={609--633},
  year={1993},
  publisher={Cambridge University Press}
}

@article{deschamps2009dynamics,
  title={Dynamics of a vesicle in general flow},
  author={Deschamps, Julien and Kantsler, Vasiliy and Segre, Enrico and Steinberg, Victor},
  journal={Proceedings of the National Academy of Sciences},
  volume={106},
  number={28},
  pages={11444--11447},
  year={2009},
  publisher={National Academy of Sciences}
}

@article{chang1993experimentalb,
  title={Experimental studies of the deformation of a synthetic capsule in extensional flow},
  author={Chang, Keun-Shik and Olbricht, William L},
  journal={Journal of Fluid Mechanics},
  volume={250},
  pages={587--608},
  year={1993},
  publisher={Cambridge University Press}
}

@article{xie2017interfacial,
  title={Interfacial rheological properties of self-assembling biopolymer microcapsules},
  author={Xie, Kaili and De Loubens, Cl{\'e}ment and Dubreuil, Fr{\'e}d{\'e}ric and Gunes, Deniz Z and Jaeger, Marc and Leonetti, Marc},
  journal={Soft matter},
  volume={13},
  number={36},
  pages={6208--6217},
  year={2017},
  publisher={The Royal Society of Chemistry}
}

@article{chen2023robust,
  title={Robust fabrication of ultra-soft tunable PDMS microcapsules as a biomimetic model for red blood cells},
  author={Chen, Qi and Singh, Naval and Schirrmann, Kerstin and Zhou, Qi and Chernyavsky, Igor L and Juel, Anne},
  journal={Soft Matter},
  volume={19},
  number={28},
  pages={5249--5261},
  year={2023},
  publisher={Royal Society of Chemistry}
}

@article{de2014mechanical,
  title={Mechanical characterization of cross-linked serum albumin microcapsules},
  author={De Loubens, Cl{\'e}ment and Deschamps, Julien and Georgelin, Marc and Charrier, Anne and Edwards-Levy, Florence and Leonetti, Marc},
  journal={Soft matter},
  volume={10},
  number={25},
  pages={4561--4568},
  year={2014},
  publisher={Royal Society of Chemistry}
}

@article{barthes2002effect,
  title={Effect of constitutive laws for two-dimensional membranes on flow-induced capsule deformation},
  author={Barthes-Biesel, Dominique and Diaz, Anna and Dhenin, Emmanuelle},
  journal={Journal of Fluid Mechanics},
  volume={460},
  pages={211--222},
  year={2002},
  publisher={Cambridge University Press}
}

@article{barthes1981time,
  title={The time-dependent deformation of a capsule freely suspended in a linear shear flow},
  author={Barthes-Biesel, Dominique and Rallison, JM},
  journal={Journal of Fluid Mechanics},
  volume={113},
  pages={251--267},
  year={1981},
  publisher={Cambridge University Press}
}

@book{lamb1924hydrodynamics,
  title={Hydrodynamics},
  author={Lamb, Horace},
  year={1924},
  publisher={University Press}
}

@book{frankel1968motion,
  title={On the motion of small particles freely-suspended in laminar shearing flows},
  author={Frankel, Neil Arnold},
  year={1968},
  publisher={Stanford University}
}

\end{document}